# Transonic Dislocation Propagation in Diamond

Kento Katagiri[1,2,3,4,5]*, Tatiana Pikuz[6], Lichao Fang[3,4,5], Bruno Albertazzi[7], Shunsuke Egashira[2], Yuichi Inubushi[8,9], Genki Kamimura[1], Ryosuke Kodama[1,2,6], Michel Koenig[1,7], Bernard Kozioziemski[10], Gooru Masaoka[1], Kohei Miyanishi[9], Hirotaka Nakamura[1], Masato Ota[2], Gabriel Rigon[11], Youichi Sakawa[2], Takayoshi Sano[2], Frank Schoofs[12], Zoe J. Smith[13], Keiichi Sueda[9], Tadashi Togashi[8,9], Tommaso Vinci[7], Yifan Wang[3,4,5], Makina Yabashi[8,9], Toshinori Yabuuchi[8,9], Leora E. Dresselhaus-Marais[3,4,5], and Norimasa Ozaki[1,2]

**Affiliations:**

[1]Graduate School of Engineering, Osaka University; Suita, 565-0871, Japan.
[2]Institute of Laser Engineering, Osaka University; Suita, 565-0871, Japan.
[3]Department of Materials Science & Engineering, Stanford University; Stanford, 94305, USA.
[4]SLAC National Accelerator Laboratory; Menlo Park, 94025, USA.
[5]PULSE Institute, Stanford University, Stanford, 94305, California, USA
[6]Institute for Open and Transdisciplinary Research in Initiatives, Osaka University; Suita, 565-0871, Japan.
[7]LULI, CNRS, CEA, Ecole Polytechnique, UPMC, Univ Paris 06: Sorbonne Universites, Institut Polytechnique de Paris; Palaiseau, F-91128, France.
[8]Japan Synchrotron Radiation Research Institute; Sayo, 679-5198, Japan.
[9]RIKEN SPring-8 Center; Sayo, 679-5148, Japan.
[10]Lawrence Livermore National Laboratory; Livermore, 94550, USA.
[11]Department of Physics, Nagoya University; Nagoya, 464-8602, Japan.
[12]United Kingdom Atomic Energy Authority, Culham Science Centre; Abingdon, OX14 3DB, United Kingdom.
[13]Department of Applied Physics, Stanford University; Stanford, 94305, USA.

*Corresponding author. Email: kentok@stanford.edu

**Abstract:** The motion of line defects (dislocations) has been studied for over 60 years but the maximum speed at which they can move is unresolved. Recent models and atomistic simulations predict the existence of a limiting velocity of dislocation motions between the transonic and subsonic ranges at which the self-energy of dislocation diverges, though they do not deny the possibility of the transonic dislocations. We use femtosecond x-ray radiography to track ultrafast dislocation motion in shock-compressed single-crystal diamond. By visualizing stacking faults extending faster than the slowest sound wave speed of diamond, we show the evidence of partial dislocations at their leading edge moving transonically. Understanding the upper limit of dislocation mobility in crystals is essential to accurately model, predict, and control the mechanical properties of materials under extreme conditions.

**One-Sentence Summary:** Femtosecond x-ray imaging reveals shock-induced dislocation motion in diamond can exceed the sound speed barrier.

**Main Text:**

Motion of dislocations inside a material caused by external stress is related to the material's mechanical properties and its deformation dynamics (*1*,*2*). When a ductile material is stressed, dislocations inside the material move to locally accommodate that force, resulting in its plasticity (*3*). Though the ductility common to metals is usually absent in brittle materials like diamond, even brittle materials can exhibit ductility under some types of extreme conditions like shock-induced high strain-rate deformation (*4-6*). While the basic mechanisms of dislocation-mediated plasticity are sometimes invariant to the rate of strain in the material as it responds to external stresses (*7*), numerous studies have observed cases that show high rate sensitivity (*8*).

At the highest strain rates, dislocations move during deformation at velocities that approach the sound speeds of the material. Dislocation theory predicts that the self-energy and stress of dislocations diverge when a dislocation approaches a limiting (or critical) velocity in a given crystal (*9-12*). This implies that dislocations are forbidden to travel at those limiting velocities. While the limiting dislocation velocities in an isotropic crystal coincides with the longitudinal and transverse sound speeds (*13*,*14*), anisotropic single-crystals like diamond have three sound speeds $c_1 \geqq c_2 \geqq c_3$ which are of the same order but do not always coincides with the limiting velocities (*10-12*,*15-18*). The slowest sound speed, $c_3$, defines the separation between subsonic and transonic speed regimes.

To date, transonic or supersonic (i.e., faster than $c_3$ and $c_1$, respectively) dislocation motion in a real crystal has not been observed experimentally. The only reported experimental evidence of dislocations moving faster than the slowest limiting velocity was in a plasma crystal (*19*). By contrast, numerous theory and molecular dynamics (MD) simulation studies have predicted the existence of transonic or even supersonic dislocation motion, indicating that the limiting velocities should not be the upper limit of dislocation motions (*9*,*20-26*). Gumbsch and Gao (*9*) used atomistic simulations to observe ultrafast dislocation motion in tungsten. Their simulations showed that the stable motion of dislocation is possible for both transonic and supersonic speeds, but only for dislocations created above the limiting speed, thus avoiding the need to accelerate across the infinite-energy limiting velocity.

While creating dislocations at such high speeds is relatively straightforward for MD simulations, experiments that measure dislocations have not been able to access those rapidly-driven conditions. With $>10^7$ $s^{-1}$ strain rates, the shock compression technique offers a unique system to study high-velocity dislocations because the energy discontinuity at the shock wavefront can create dislocations initially moving faster than the limiting velocity (*15*,*27*). Here we present experimental results that show shock-induced dislocation motions moving transonically in single-crystal diamond, using femtosecond x-ray radiography. We discuss our interpretation and assignment of the relevant image features and include a discussion of how the dislocation speeds and related plasticity indicate radiation and drag that imply a relationship between the microstructure and bulk elastic-plastic behaviors. Understanding the speed of the fastest dislocation motion is required to accurately predict and control the dislocation dynamics and plasticity of deformed solids, which can be modeled by discrete dislocation dynamics (*28,29*). Such ultrafast motion of dislocations strongly affects the mechanical responses of materials in ways that are essential across numerous applications.

**Visualization of stacking faults**

Our experiments were conducted using the X-ray Free Electron Laser (XFEL) at SPring-8 Angstrom Compact Free Electron Laser (SACLA). At SACLA's Experimental Hutch 5, a high-intensity nanosecond-pulsed optical laser is aligned to the same spatial position as the femtosecond-pulsed XFEL, and its timing is synchronized to perform *in-situ* x-ray measurements of laser-shocked materials (*30*). We use type IIa single crystals of diamond synthesized by chemical vapor deposition. Our diamonds were cut along two different orientations to enable our shock to propagate along the [100] and [110] directions, to examine the orientational difference in the plasticity mediated by shock-induced defects (see Fig. S9 for [111] shock data). We estimate the peak shock stresses applied to the diamond samples to be 184 ± 16 GPa and 92 ± 15 GPa for [100] and [110] shock directions, respectively (*18*). These shock stresses are high enough to make diamond yield (*5*), generating strong shear stresses that initiate failure mechanisms. When diamond yields during shock compression, the shock wave splits into two wavefronts, leading with an elastic wavefront that is followed by the plastic wavefront in which dislocation propagations occur.

While several x-ray imaging techniques are used for in-situ measurements (*31-34*), x-ray radiography is one of the original ones used to non-destructively visualize structures in bulk materials (*35*). X-ray radiography spatially maps the x-ray intensity transmitted through a material of interest, giving image contrast that is usually based on the absorption by the sample. The contrast we observe in this work is mainly due to the scattering of the x-rays rather than absorption; we note that despite the similarity in contrast mechanism, our radiography setup is distinct from the phase contrast imaging setups used in other XFEL shock experiments (*32*). To perform the in-situ x-ray radiography, we use an unfocused and single-pulsed XFEL beam with a photon energy of 7.0 keV and pulse duration of ~8 fs to illuminate the diamond along the axis perpendicular to the shock propagation direction (Fig. 1). We place a lithium fluoride (LiF) crystal downstream of the sample along the XFEL beam as a vacuum-compatible imaging detector that offers a $>10^6$ dynamic range and ~1 μm spatial resolution over a wide field of view (*36*). The distance between the diamond sample and the front surface of the LiF crystal is 112 mm.

Our x-ray radiography for a diamond shocked along the [110] direction captures two shock wavefronts, corresponding to a preceding elastic wave and a following plastic wave, traversing the diamond sample from the bottom to the top of the image (Fig. 1). Behind the plastic wavefront, light and dark bands appear diagonal to the shock direction. In all cases, the two characteristic vectors of the shock wave and the XFEL imaging beam's direction indicate the angles of the linear banded features appear parallel to the {111} planes of the crystal. When the {111} planes are set not along the XFEL's probing direction (i.e., the sample was placed 90° rotated about the shock propagation axis), these banded features appear significantly fainter (Fig. S10). This indicates that the image contrast of the banded features in our x-ray radiography is strongest when the x-ray beam direction integrates along a plane perpendicular to the 2D features formed in the diamond.

We use Fourier filtering methods to emphasize the linear image features (Fig. 2C, F). These banded features in the [110] shock images appear thicker and darker than those seen in the [100] shock images. As described in the next section, the dislocation propagation for the [110] shock is faster than that for the [100] shock, which potentially affects the thickness and darkness of the stacking faults as a faster dislocation propagation results in faster macroscopic plastic flow (*38*) causing more damage to the material. We measure the angles between the linear features and the diamond surface using the Hough transform on the Fourier-filtered images (*18*). The observed angles of the 16 ns images are 56 ± 3 and 32 ± 3° for [100] and [110] shocks, respectively. While the value for [100] agrees with the angles between the diamond surface and the {111} plane of the undistorted diamond lattice which is 55° (Fig. 2B), the observed angle for [110] shows a relatively larger deviation from the angle of the undistorted diamond lattice which is 35° (Fig. 2E). Although the deviation is within the estimated error, we believe this deviation suggests that the diamond shocked along [110] experiencing a plastic deformation with some distortion caused by the uniaxial shock compression. We observed bands along non-{111} directions only in the [100] shock image collected at 16 ns delay (Fig. 2C, orange arrows). The angular offset may indicate shock-induced twinning or slip along another crystallographic plane, though we cannot exclude the possibility of the diamond crystal having an undesirable formation of a rotated grain during the synthesis.

We interpret these deformation-induced bands as accumulated stacking faults (*39*) forming along the {111} planes. The {111} plane in diamond is known to be the dominant slip plane for dislocations. Previous shock-loading experiments in other brittle materials have reported the occurrence of shock-induced amorphous bands formed diagonally to the shock wavefront (*40*-*44*), but the estimated temperature in the shocked diamond in our work is less than 20% of its melting temperature (*18*), which we interpret as too low for amorphization. The possibility of phase transformations (*45*,*46*) is also negligible because of the low stress applied to the diamonds.

**Transonic dislocation propagation**

Elastic theory predicts that transonic dislocations can be generated at discontinuities in energy or displacement within a material like a shock wavefront (*15*,*27*). Under these conditions, a full dislocation would disintegrate into partial dislocations, leaving a stacking fault between them (*47*,*48*). As the edges of a stacking fault are defined by partial dislocations (*49*,*50*), the leading edge of the stacking-fault extension we observe gives the dislocation velocity in our experiment. In our images, the stacking faults formed in the diamonds appear as discontinuous lines, indicating the dislocations travel with the plastic wavefront, originating from the laser-irradiated side of the diamond surface. A closer look at the [110] shock images (Fig. 1B) shows the absence of these bands at the center of the plastic wavefront. This observation is because most of the dislocation motions behind the plastic wavefront are propagated from the limited area of the diamond surface that is shocked by the drive laser ($\varphi \sim 260$ µm). This indicates that the number of dislocations freshly created at the propagating plastic shock front is much less than those propagating from the diamond surface.

We plot the length of the stacking faults we observe in our images ($x_d$) as a function of time delay after the shock initiation ($t$) (Fig. 3A). Then we quantify the velocity of the partial dislocations ($v_d$) leading that stacking-fault extension by fitting the points to a line with $v_d = dx_d/dt$ from the measured $x_d$ - $t$ relationship. We show the obtained dislocation velocities (Fig. 3B) along with the sound wave velocities ($c_1$, $c_2$, and $c_3$) of diamond at various densities (*18*) calculated from the pressure-dependent elastic constants at 1,100 K from (*51*). The dislocation velocities we observe for the [100] and [110] shock directions both lie in the transonic regime (i.e., between $c_1$ and $c_3$). We find that the velocity of dislocations in the [110] shocked crystals is faster than those of the [100] crystal. While the plastic wave velocities differ only slightly, the angle between the shock direction and dislocation propagation direction is larger for the [110] shock than the [100] shock, making the dislocation propagation distance longer (and thus faster) for the [110] to catch up with the plastic shock wavefront.

The existence of supersonic dislocation has been observed in several MD simulations though some of the early theoretical studies predicted it to be impossible (*27*,*52*). Because the shock-induced dislocation motions we observe in diamond are traveling with the plastic shock wavefront, applying a higher shock stress driving a faster plastic shock wave can potentially drive even faster dislocation motion. This will allow one to experimentally investigate the possible existence of supersonic dislocations in a real crystal.

**Radiation from transonic dislocations**

A close look at the elastically compressed volume immediately preceding the plastic wavefront reveals the presence of multiple "elastic pulses" (Fig. 1B, yellow dotted line). These pulses are referred to as radiation (or phonon radiation) in dislocation theory (*27*,*49*). While the emission of radiation from transonically moving dislocations has been predicted by dislocation theory and observed in MD simulations (*9*,*23*,*24*), a dearth of validation methodology has precluded experimental observation of the radiation.

While we observe the radiation feature in our images of the [110] shock (Fig. 2D), our images for the [100] (Fig. 2A) and [111] shock directions (Fig. S9) show less clear signs of radiation. The different amounts of radiation we observe for shocks along different directions with different dislocation velocities should be due to the strong velocity dependence on the energy dissipation in radiations which has also been predicted by the theory and simulations (*9*,*53*). Because the radiation from the transonically propagating dislocation cores exerts a negative driving force that adds drag to the dislocation motion, large external stress is required to drive the stable motion of transonic dislocations without deceleration (*9*). The use of strong shock waves can provide stress high enough to overcome the drag by radiation to achieve stable transonic motion of dislocations.

Although the existence of the radiation from transonically propagating dislocation cores and its role in the energy dissipation have been extensively studied, its effect on the shock-induced deformation dynamics remains largely unknown. A few studies have used MD simulations of shocked single-crystal aluminum (*54*,*55*) to predict shock-induced elastic-plastic deformation mechanisms wherein the preceding elastic volume was disturbed by the plasticity that followed. Their observations were explained by the elastic pulses emitted from the dislocations at the plastic wavefront, consistent with the configuration we observe in this diamond study. A recent billion-atom MD simulation of single-crystal diamond shocked along the [110] axis revealed complex defect structures at a peak stress slightly higher than 400 GPa (*56*). Such dislocation-driven elements of shock deformation dynamics might explain the complex elastic-plastic correlations in shocked diamond (*57*) and the anomalous elastic behavior in shocked silicon (*58*) observed in previous experiments by using a conventional velocimetry.

In conclusion, our *in-situ* x-ray radiography using one of the world's brightest x-ray lasers shows experimental evidence of transonic dislocation motion in diamond, leading to the formation of numerous stacking faults that span the plastically deformed volume. The microscopic dislocation motions emitting radiation might affect the macroscopic elastic-plastic deformation dynamics. Our experimental results showing transonic dislocation motion now open key new opportunities to refine models for insights into the ultrafast deformation behaviors at these extreme conditions. The newly refined models at the highest strain rates will have a marked influence on numerous fields including ultrafast fracture in structural materials (*59*-*61*), prediction and analysis of earthquake ruptures (*62*-*65*), precise manufacturing processes (*66*), and functionalities in electrochemical applications (*67*).

**Acknowledgments:** We thank H. Sumiya, Y. Seto, T. Okuchi, and T. Duong for discussions and technical staff of SACLA for their support during the experiments. The experiments were performed at BL3 of SACLA with the approval of the Japan Synchrotron Radiation Research Institute (proposal nos. 2021A8004 & 2021B8067). The high-power nanosecond laser at SACLA was installed by Institute of Laser Engineering of Osaka University with the corporation of Hamamatsu Photonics. The installation of Diffractive Optical Elements (DOE) to improve the smoothness of the drive laser pattern was supported by the SACLA Basic Development Program. The current affiliation of G.R. is Massachusetts Institute of Technology, USA.

**Funding:**

MEXT Quantum Leap Flagship Program (MEXT Q-LEAP) Grant No. JPMXS0118067246

Japan Society for the Promotion of Science (JSPS) KAKENHI Grants Nos. 21J10604 and 20H00139

JSPS Core-to-Core Program B: Asia-Africa Science Platforms Grant No. JPJSCCB20190003

Genesis Research Institute, Inc. (Konpon-ken, Toyota)

K.K. acknowledges support from Yamada Science Foundation

B.K. acknowledges support from U.S. Department of Energy by Lawrence Livermore National Laboratory under Contract DE-AC52-07NA27344

K.K. and L.D.M. acknowledge support from Air Force Office of Scientific Research "FA9550-23-1-0347", managed by Dr Chiping Li and Capt. Derek Barbee (PhD).

Y.W. was supported by the Stanford Energy Postdoctoral Fellowship through contributions from the Precourt Institute for Energy, Bits & Watts Initiative, StorageX Initiative, and TomKat Center for Sustainable Energy.

**Author contributions:**

Conceptualization: KK, NO

Methodology: KK, TP, LEDM

Resources: FS

Investigation: KK, TP, BA, SE, YI, GK, MK, GM, KM, HN, MO, GR, YS, TS, ZJS, KS, TT, TV, MY, TY, LEDM, NO

Visualization: KK, LF, ZJS

Formal Analysis: KK, TP, LF, BK, YW

Funding acquisition: KK, LEDM, NO, YW

Project administration: NO

Supervision: RK, LEDM

Writing – original draft: KK, LEDM

Writing – review & editing: All authors

**Competing interests:** Authors declare that they have no competing interests.

**Data and materials availability:** All data are available in the main text or the supplementary materials.

**Supplementary Materials**

Materials and Methods

Supplementary Text

Figs. S1 to S12

Tables S1

References (*68*–*78*)

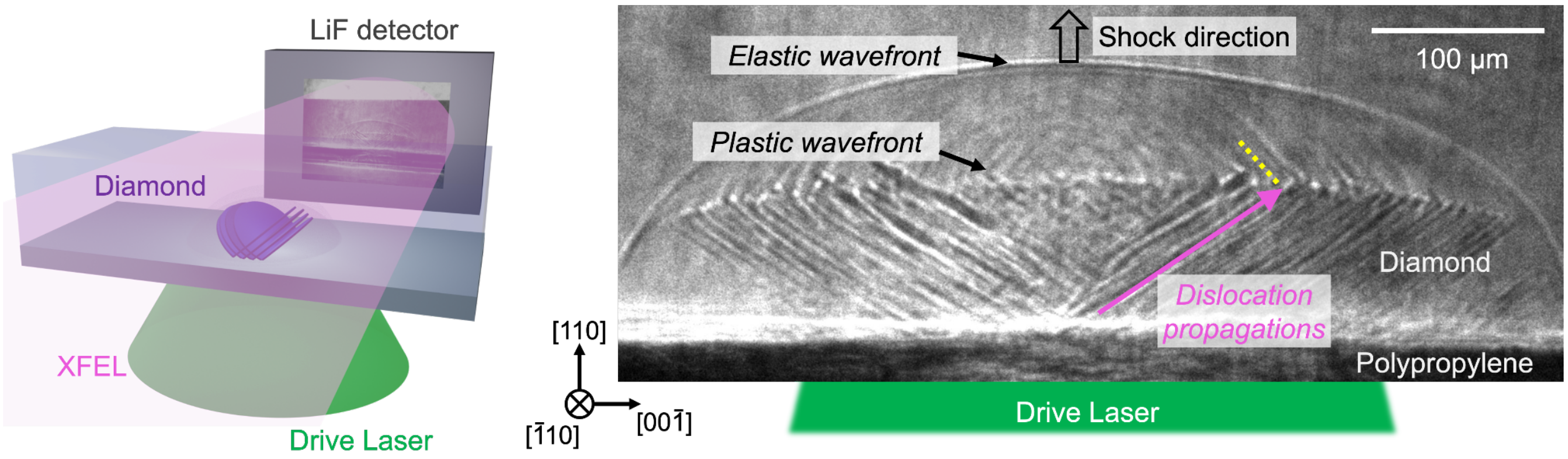

**Fig. 1. Femtosecond x-ray radiography on shocked diamond.** (**A**) The diamond sample is stressed by shock waves driven by optical laser irradiation. *In-situ* x-ray radiography is measured by using a femtosecond XFEL pulse, irradiating the sample perpendicular to the propagation axis of the shock waves. A lithium fluoride (LiF) crystal is placed at the XFEL downstream to collect the spatial intensity distribution of the transmitted x-rays. (**B**) A representative image we capture by using the LiF crystal visualizing the elastic-plastic shock wavefronts propagating along the shock direction. Behind the plastic wavefront, stacking faults appear as dark and light bands, indicating the existence of partial dislocations traveling with the plastic shock wavefront (magenta allow). The yellow dotted line is to guide eyes on the radiation.

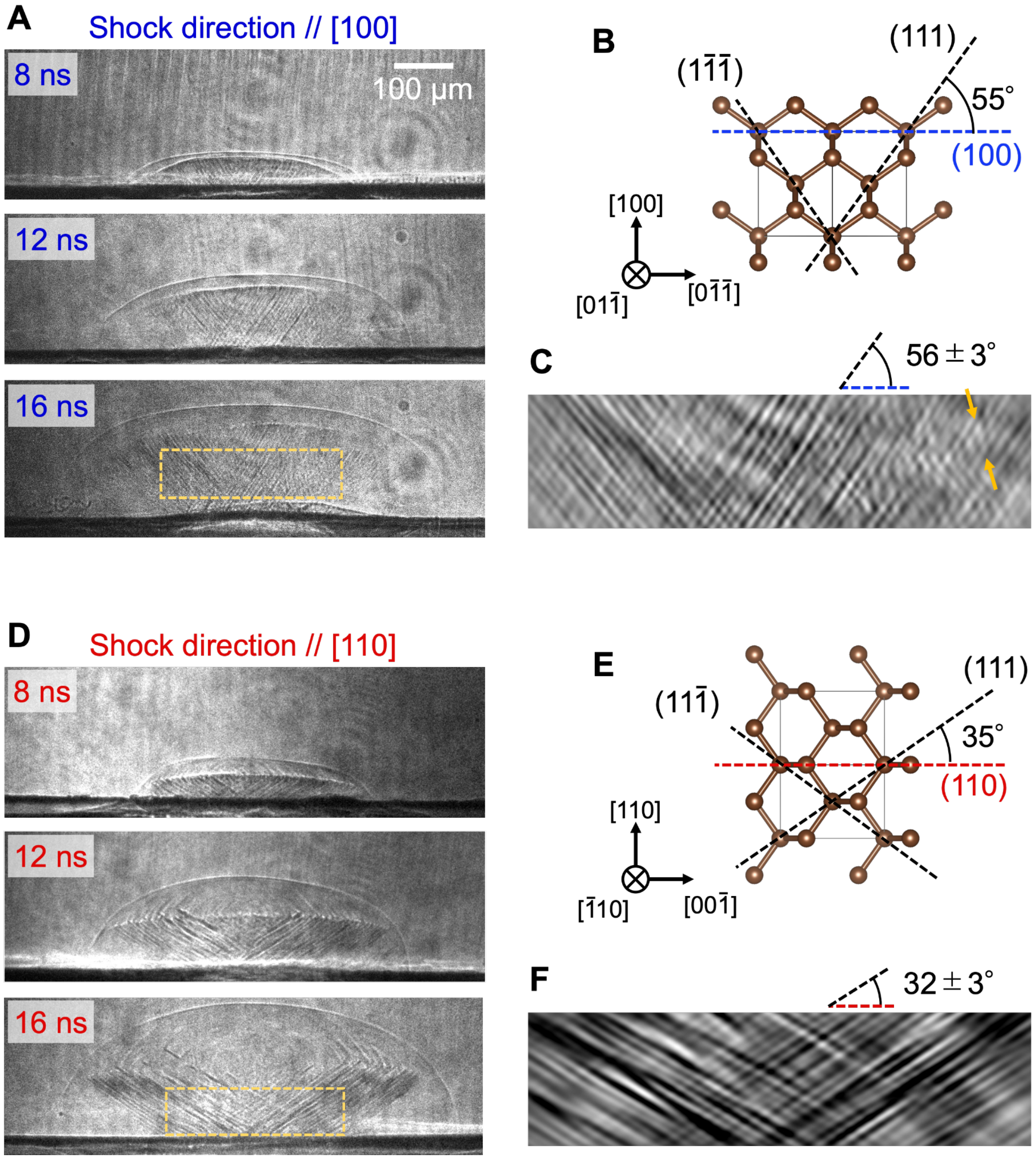

**Fig. 2. Visualization of stacking faults.** (**A**) X-ray radiography images of the diamond shocked along the [100] direction. The numbers on each image indicate the XFEL probe timing relative to the drive laser irradiation. The vertical lines that appear across the entire field of view are the texture on the unpolished lateral surfaces of the diamond. The two large circles with concentric scattering patterns are caused by scattering from the debris on a beryllium window upstream of the XFEL. (**B**) Undistorted diamond lattice structures (*37*) with the corresponding plane orientations. (**C**) FFT-filtered image showing the stacking faults observed in the 16 ns shot. The corresponding area is indicated as the yellow dashed rectangle in the image on (A). The orange arrows are to guide eyes to one of the bands formed nonparallel to the diamond {111} planes (see text). (**D**) X-ray radiography images when diamonds are shocked along [110] direction. The 12 ns image is identical to the one shown in Fig. 1B. (**E**) Undistorted lattice structure showing

the plane orientations. (**F**) FFT-filtered image of the [110] shock at 16 ns. The shown errors on the angles noted in (C) and (F) represent 1σ of multiple measurements (*18*).

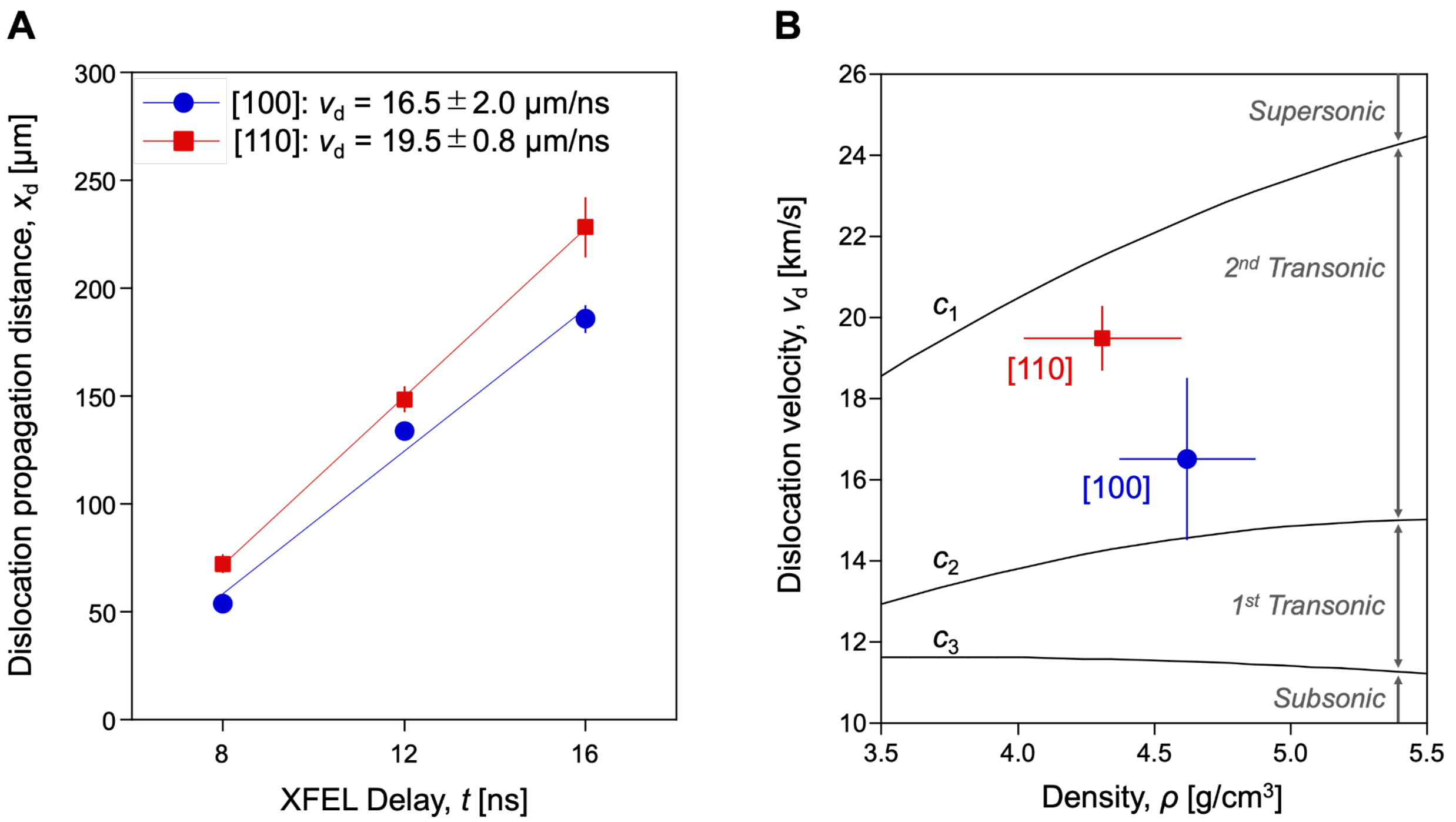

**Fig. 3. Transonic dislocation motion in shocked diamonds.** (**A**) Measured dislocation propagation distances in diamond ($x_d$) as a function of the XFEL delay ($t$). The $x_d = 0$ μm denotes the interface between the ablator and diamond, while $t = 0$ ns denotes the time that the laser hit the ablator. The blue circles and red squares represent our experimental results for the [100] and [110] shock directions, respectively. The velocities of the dislocation propagation ($v_d$) noted are obtained from the slope of the linear fits (solid lines). The error bars on each plot represent 1σ of multiple measurements. The shot-by-shot fluctuation of drive laser intensity is not included in the shown error bars on $x_d$. The errors on $v_d$ are evaluated based on the 1σ of the fitting and the errors on each plot. (**B**) $v_d$ versus the diamond's material density ($\rho$) at the shocked states. Black curves are the calculated $c_1$, $c_2$, and $c_3$ of diamond propagating along [110] direction (*18*). The error bars on $\rho$ are propagated from the errors on the plastic shock wave velocities (*18*).

# Supplementary Materials for

## Transonic Dislocation Propagation in Diamond

Kento Katagiri[1,2,3,4]*, Tatiana Pikuz[5], Lichao Fang[3,4], Bruno Albertazzi[6], Shunsuke Egashira[2], Yuichi Inubushi[7,8], Genki Kamimura[1], Ryosuke Kodama[1,2], Michel Koenig[1,6], Bernard Kozioziemski[9], Gooru Masaoka[1], Kohei Miyanishi[8], Hirotaka Nakamura[1], Masato Ota[2], Gabriel Rigon[10], Youichi Sakawa[2], Takayoshi Sano[2], Frank Schoofs[11], Zoe J. Smith[3], Keiichi Sueda[8], Tadashi Togashi[7,8], Tommaso Vinci[6], Yifan Wang[3,4], Makina Yabashi[7,8], Toshinori Yabuuchi[7,8], Leora E. Dresselhaus-Marais[3,4], and Norimasa Ozaki[1,2]

Corresponding author: kentok@stanford.edu

**The PDF file includes:**

# Materials and Methods

## Laser-shock compression and XFEL probe

Our experiments are performed by using the X-ray Free Electron Laser (XFEL) (*68*) at SPring-8 Angstrom Compact Free Electron Laser (SACLA) (*30*). Each shock target consists of a single crystal diamond sample and a 50-μm thick polypropylene ablator. The ablator is attached to the diamond sample by using UV-cured resin (ThreeBond Product No. 3042) and the measured thickness of the resin layer is ~1 μm. The thickness of the diamond sample along the shock propagation direction varies between 200-300 μm for different orientations, while the thickness along the x-ray irradiation direction is 500 μm for all orientations. The diamond surface contacting the ablator is polished. All of the diamonds we use in this work are type IIa single crystalline diamond synthesized by using the chemical vapor deposition (CVD) method, used as-is after purchasing from EDP Corporation (Japan).

A flat-top optical laser pulse with a 5-ns pulse duration is irradiated on the target, focused to a 260-μm diameter spot size at the sample. Diffractive Optical Elements (DOE) was used to smooth its spatial mode. The on-target laser energies are 17.4 ± 1.4 and 16.9 ± 1.2 J for the [100] and [110] orientations, respectively.

To perform the *in-situ* x-ray radiography, an XFEL pulse with a photon energy of 7.0 keV and pulse duration of ~8 fs is irradiated on the diamond target perpendicular to the shock propagation direction. The XFEL beam is unfocused and has a beam divergence of ~2 μrad. The XFEL beam diameter is ~800 μm at the target assembly's position. The distance between the laser-shocked portion of the diamond sample and the front surface of the LiF crystal is 112 mm.

Diamonds with three different orientations along the shock propagation directions: [100], [110], and [111] are used in this experiment. Though the data for diamond shocked along [100] and [110] are presented in the main text, the data for [111] is summarized in the Supplementary Text as we could not obtain the dislocation velocity for the [111] orientation.

## X-ray imaging using a LiF crystal

When a lithium fluoride (LiF) crystal is exposed to high-intensity x-rays, it forms color centers with a local density that scales with the local x-ray intensity. Color centers are point defects in ionic crystals in which holes and electrons are trapped, producing localized features that can be observed with visible light. The spatial distributions of the x-ray intensities irradiated to the LiF crystal can be read out by using a confocal fluorescence microscope that measures the visible-light fluorescence generated from these color centers. Hence, LiF crystals work as a vacuum-compatible x-ray imaging detector with a high spatial resolution as previously demonstrated in (*36,69-73*). In this work, we use a confocal fluorescence microscope (Nikon C2) with an excitation wavelength of 445 nm for the read-out process. The magnification of the confocal microscope during the read-out process is 20x (for the images on Figs. 1B and 2C&F) and 10x (other images) unless otherwise noted. The spatial resolution of the x-ray radiography is limited by the resolution of the visible-light microscope (0.31 and 0.62 μm/pixel for 20x and 10x magnifications, respectively) and by the x-ray propagation contrast arising from the Fresnel number of the relevant image features.

**Shock wave velocity estimation**

The measured relationships between the elastic wave's travel distance in the images ($x_E$) and the XFEL probe time relative to the drive laser irradiation timing ($t$) are shown in Fig. S1. We estimate the elastic wave velocities ($D_E$) from the slope of the linear fit (i.e., $D_E = dx/dt$). The elastic wave velocities for [100], [110], and [111] orientations we determine are 17.9 ± 0.4, 20.4 ± 0.4, and 20.2 ± 0.8 km/s, respectively. These values agree with the elastic shock wave velocity of single-crystal diamond at its Hugoniot elastic limit measured by using velocimetry (*5*). The elastic shock front position for [111] at 16 ns is not determined as the shock wave has already reached the rear surface at the time we probe (Fig. S1).

For the slower plastic wave, the relationships between the measured plastic wave travel distance ($x_P$) and the XFEL probe time relative to the drive laser irradiation timing ($t$) are shown in Fig. S2. The plastic shock velocities we obtain from the slope of the linear fit of the plots shown in Fig. S2 are 14.8 ± 0.4 and 11.9 ± 0.5 km/s for [100] and [110], respectively. The velocity for [111] is not obtained as the plastic wavefront positions for 12 and 16 ns shots are not clear. This is because the separations between the elastic and plastic volumes are blurred at late times which was also seen in a previous study of shocked diamonds with a similar x-ray imaging setup (*73*). This blurring could be a result of a rarefaction wave propagated from the drive laser irradiation side, but such blurring is not seen in [100] and [110] shocks and its reasons remain unclear.

Since the duration of the drive laser pulse is 5 ns and we use a 50 μm thick polypropylene ablator, the rarefaction wave(s) should be present in the diamond at the times the XFEL was irradiated. Though sampling only three delay times for each orientation may not be sufficient to discuss the shock decay behaviors, the small divergence of the plots from the linear fits shown in Figs. S1 and S2 suggest that there are no significant decays of the shock wavefronts at least for the times up to 16 ns.

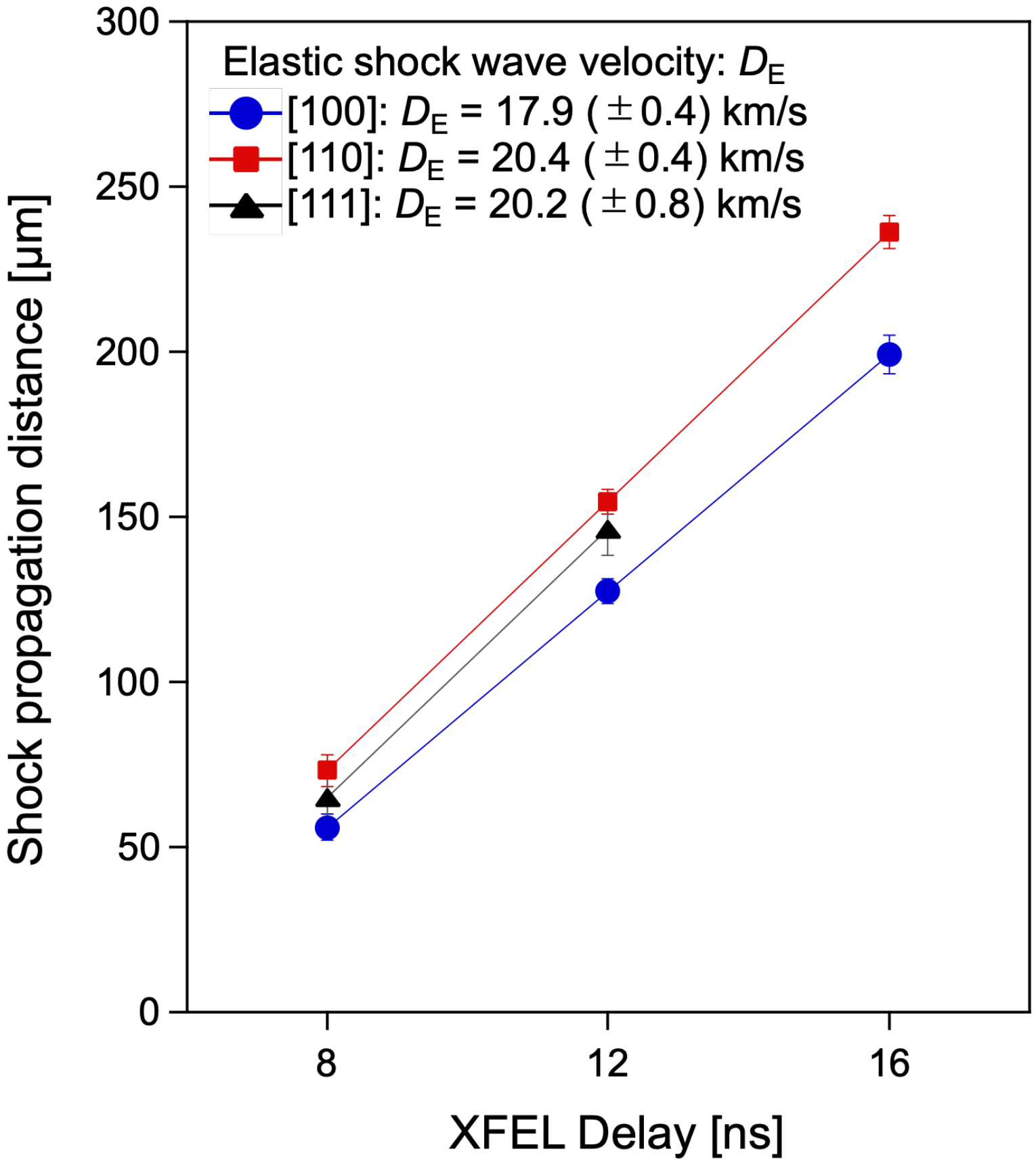

**Fig. S1.**
**Elastic shock wave propagation distances as a function of the XFEL delay time relative to the drive laser irradiation timing.** The elastic wave positions are measured separately for [100] (blue circles), [110] (red squares), and [111] (black triangles). The elastic shock velocities ($D_E$) obtained by taking the linear fits of the plots (solid lines) are noted. The 0 μm for the vertical axis denotes the interface between the ablator and diamond, while 0 ns for the horizontal axis denotes the time that the laser hit the ablator. The shown errors on each plot represent systematic errors. The noted error on $D_E$ is evaluated based on the 1σ of the fitting and the errors on each plot.

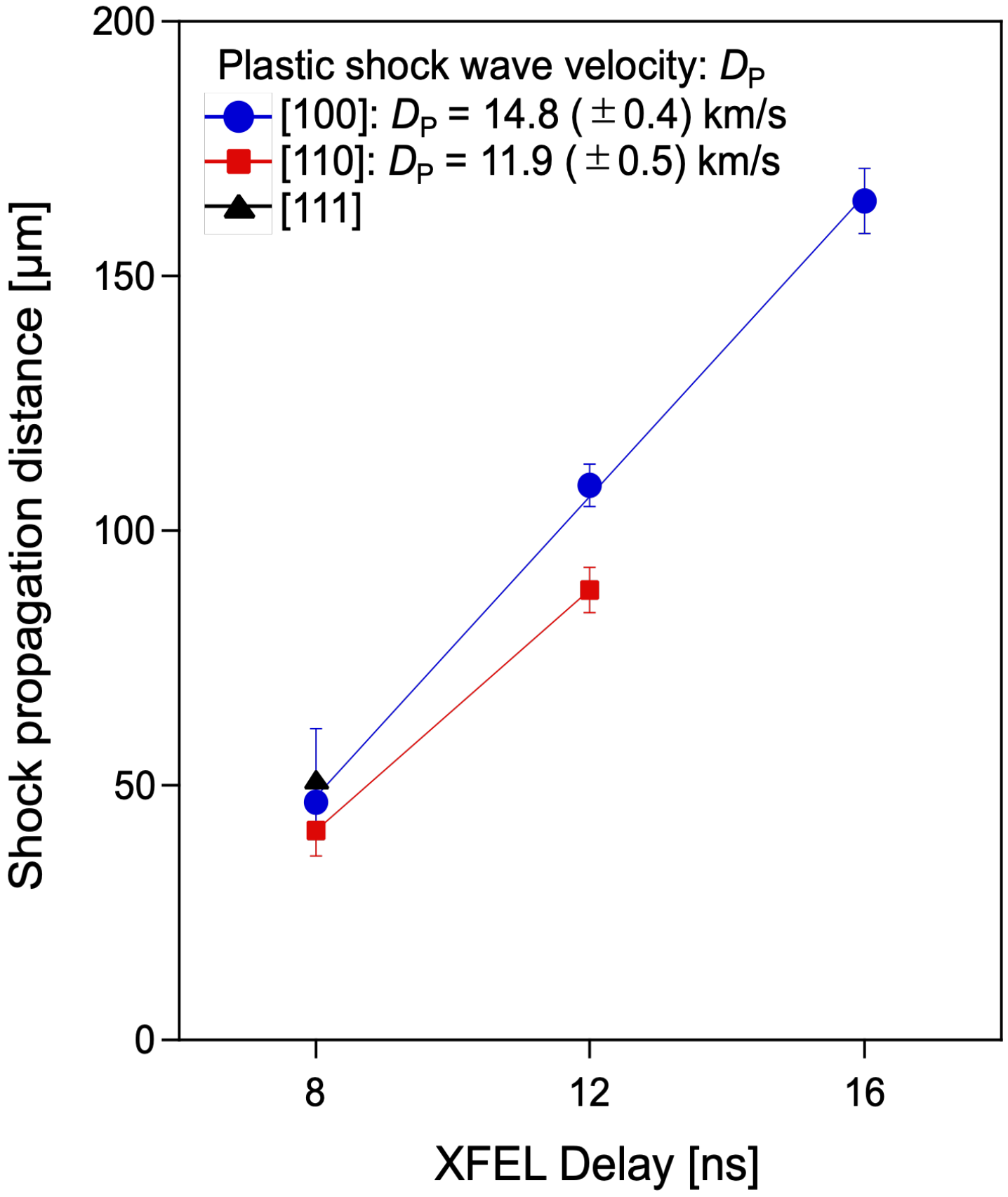

**Fig. S2.**
85 **Plastic shock wave propagation distances as a function of the XFEL delay.** The symbols and colors are the same as in Fig. S1. The shown errors on each plot represent systematic errors. The noted error on $D_P$ is evaluated based on the 1σ of the fitting and the errors on each plot.

## Estimation of the density, stress, and temperature at the shock-end states

We convert the determined plastic shock wave velocities ($D_P$) to the density, stress, and temperature of the diamond at each peak-shock state (Fig. S3) by applying the existing Hugoniot equation-of-state of diamond. The Hugoniot curves measured in laser-shock experiments (*5*) were used to estimate the shock stress and density, while the shock temperature was estimated from the stress-temperature relationship of the shocked diamond predicted by molecular dynamics simulation results from (*74*). The estimated peak shock stresses applied to the diamond samples were 184 ± 16 GPa and 92 ± 15 GPa for [100] and [110] shock directions, respectively. The difference of the stresses between [100] and [110] is larger than what we expected as the difference of the on-target laser energies were smaller (17.4 ± 1.4 and 16.9 ± 1.2 J for [100] and [110] orientations, respectively). Since the shock transit times in the ablator seem to not vary significantly between the [100] and [110] shots, the difference in the final stresses between [100] and [110] shock directions could be largely due to the difference in the shock impedances between them, which have not been measured precisely at these stress regions.

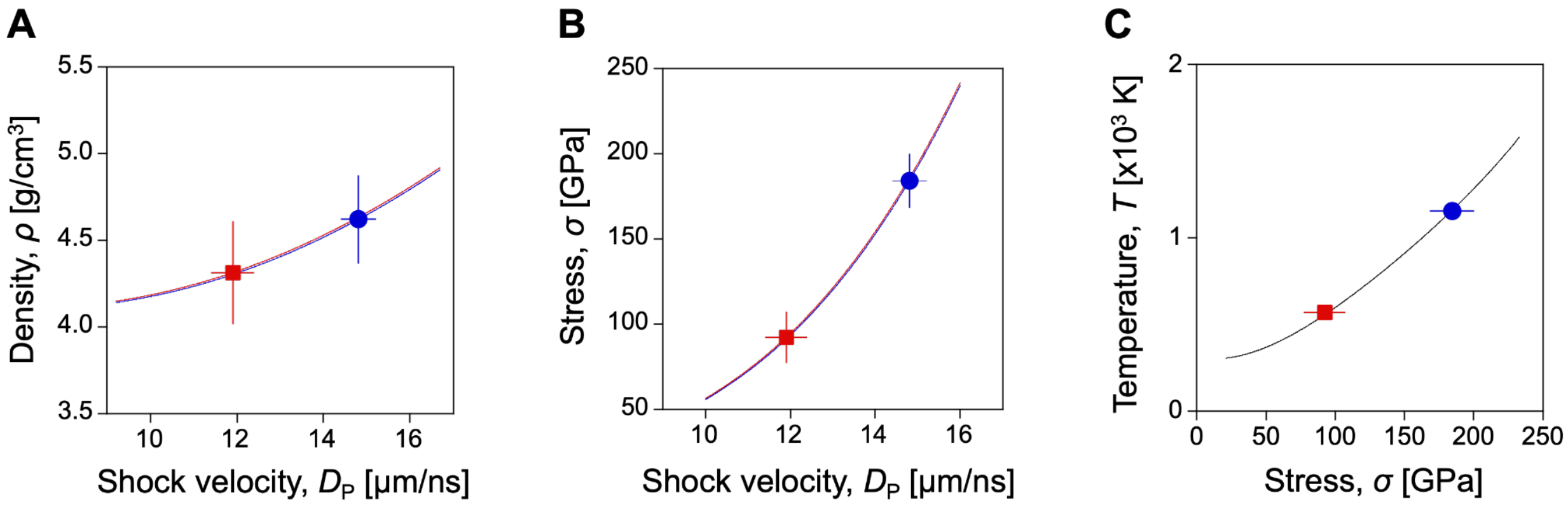

**Fig. S3.**
**Characterization of the peak-shock states in the diamond. (A)** Densities, (**B**) stresses, and (**C**) temperatures at each shock end-state of [100] (blue circles) and [110] (red squares) compressions estimated from the determined plastic shock wave velocities. The blue and red curves in A and B represent the Hugoniot curves for [100] and [110] compression, respectively (*5*). The black curve in C shows the Hugoniot temperature simulated in ref (*74*). The shown errors in densities and stresses are propagated from the errors in the plastic shock wave velocity (Fig. S2). The error bars for the temperature are not estimated.

**Density-dependent sound wave velocities of diamond**

The density-dependent sound wave velocities of diamond shown in Fig. 3B of the main text were determined using the pressure-dependent elastic constants of diamond at 1,100 K reported by Valdez et al (*51*) (Fig. S4A). The pressure-density relationship (Fig. S4C) calculated by McDonald et al (*75*) was used to convert the estimated pressure to the density (Fig. S4B). In Fig. 3b, we chose to show the [110] sound speeds since the limiting velocities coincide with the sound velocities for [110], making the discussion simpler.

The three sound speeds $c_1$, $c_2$, and $c_3$ in anisotropic solids are determined for every propagation direction by solving the eigenvalue problem (*16*):

$$\det(\hat{v} \cdot C \cdot \hat{v} - \rho v^2 \mathbb{1}) = 0$$

where $\hat{v}$ is the unit vector of the wave propagation direction ([112] for 0 deg and [110] for 90 deg), $C$ is the tensor describing the elastic constants, and $\mathbb{1}$ is the $3 \times 3$ identity matrix. The solutions for the pressure at 200 GPa ($C_{11}$ = 1883 GPa, $C_{12}$ = 678.73 GPa, $C_{44}$ = 950.34 GPa) are calculated for all the propagation directions, as shown in Fig. S5. While the velocity varies as a function of crystallographic direction, this change is relatively small and thus does not change our conclusion of transonic dislocation motion (i.e., the obtained dislocation velocities shown in Fig. 3b fall between $c_1$ and $c_2$).

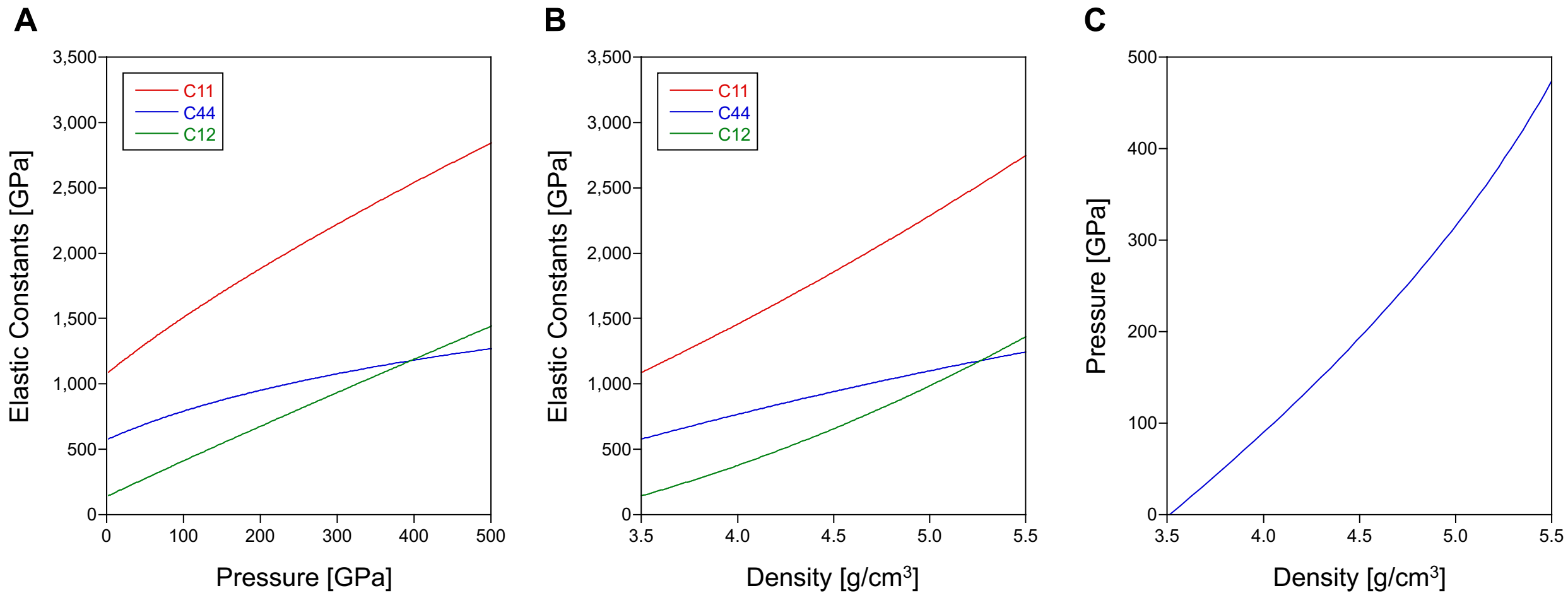

**Fig. S4.**
**Elastic constants at high pressure and density.** (**A**) The relationship between the pressure and elastic constants of diamond at 1,100 K reported in (*51*). (**B**) The density dependence of the elastic constants of diamond at 1,100 K converted from (A) using the pressure-density relationship presented in (C). (**C**) The relationship between density and hydrostatic pressure reported in (*75*).

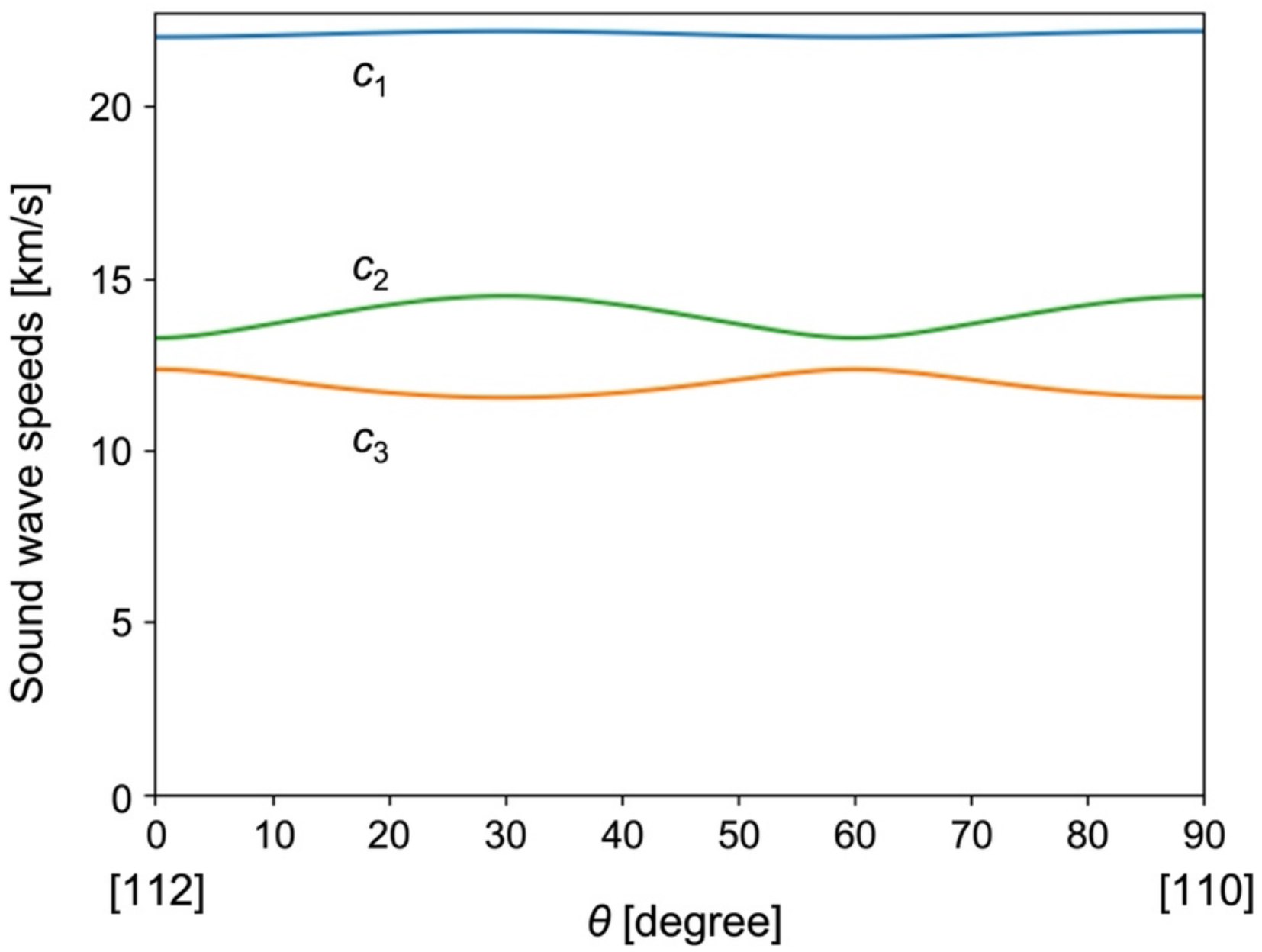

**Fig. S5.**
**The direction dependence of the sound-wave speed of diamond at 200 GPa**. The $\theta = 90$ deg corresponds to [110] direction (edge dislocation) while $\theta = 0$ deg corresponds to [112] direction (screw dislocation).

**Line detections by FFT and Hough transform**

To measure the angles between the observed stacking faults and the diamond surface, we performed line detections using a Hough transformation to capture the distribution of angles in the lines. We first apply a fast Fourier transform (FFT) filter to the x-ray images (Fig. S6) to reduce background noise and filter out other features. Then, we performed the Hough transform on the FFT images to automatically extract lines and calculate the angles of lines.

For initial FFT analysis, the full-resolution image is first manually cropped to only include the regions including the linear features appearing behind the plastic wavefront (Fig. S6A). The cropped images are then decomposed from the spatial to the frequency domain (Fig. S6B). Using a rectangle image as an input for FFT often introduces artifacts but such artifacts were filtered in the Fourier domains and thus its effect on the inverse Fourier transformed images and line detection is negligible. The Fourier spectrum calculated from FFT was filtered by setting a signal count threshold that was tailored for each image (determined by manual inspection for each image) (Fig. S6C). The inverse Fourier transform was then applied to the filtered Fourier image (Fig. S6D).

Using the Fourier-filtered images, we then perform Hough transformations to quantify the linear features. A typical result of the line detection and angle measurements of the detected lines are shown in Fig. S7. As shown in Fig. S7A, we identify and calculate the lines with angles from the FFT-filtered images with the Hough transform method. The angles of the features we detect in each x-ray radiography image of the diamond shocked along [100], [110], and [111] directions are listed in Table S1 and shown in Fig. S8.

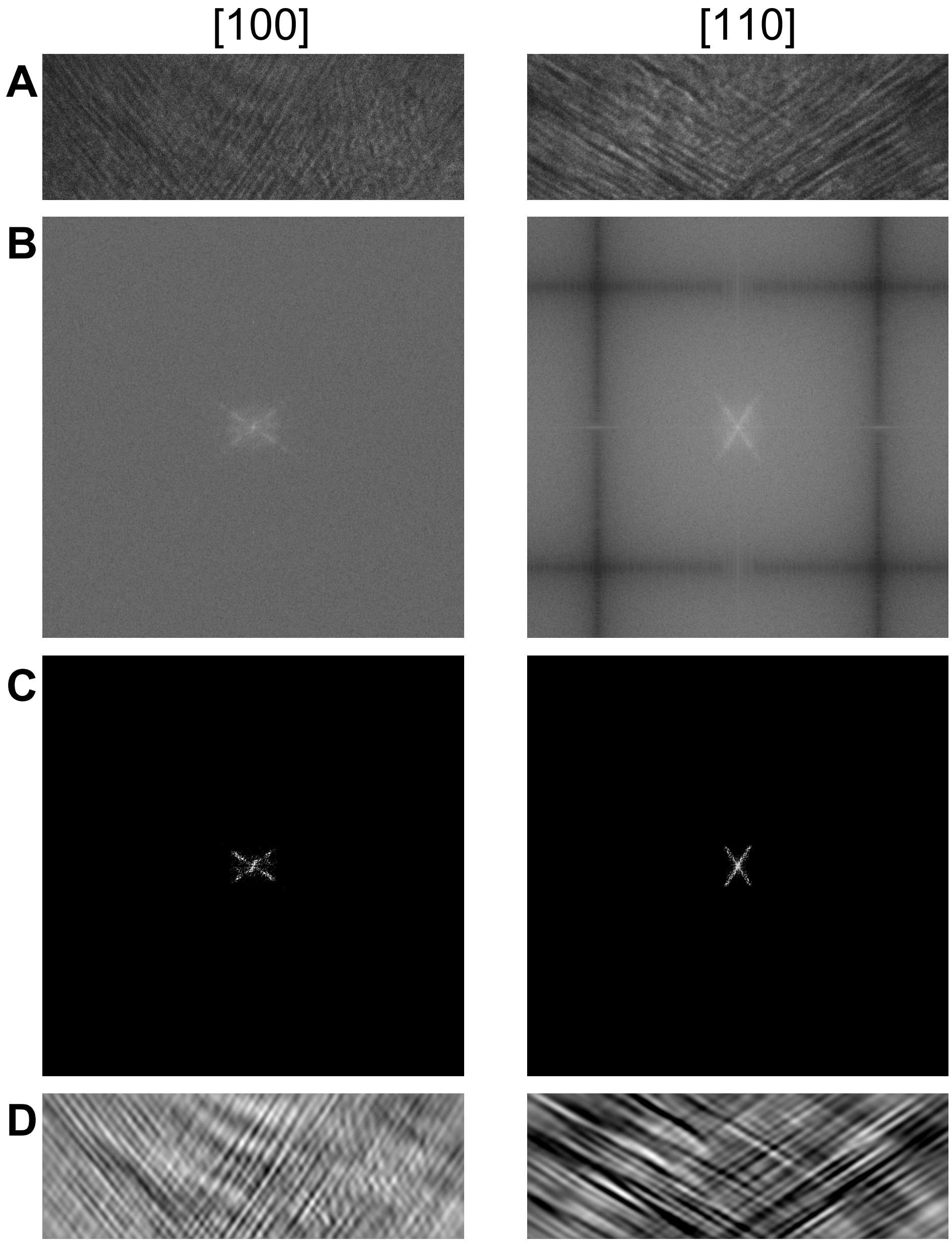

**Fig. S6.**

**Process of the Fourier filtering method applied to the images.** Two shock directions [100] (left) and [110] (right) are shown. (**A**) The cropped x-ray radiograph images. The corresponding area is indicated as the yellow dashed rectangle in the image in Fig. 2A and D. (**B**) FFT images of (A). (**C**) The signals weaker than the threshold values tailored for individual images were filtered out. (**D**) Inverse FFT images of (C) showing the enhanced linear features. These images are identical to the ones shown in Fig. 2C and F, respectively.

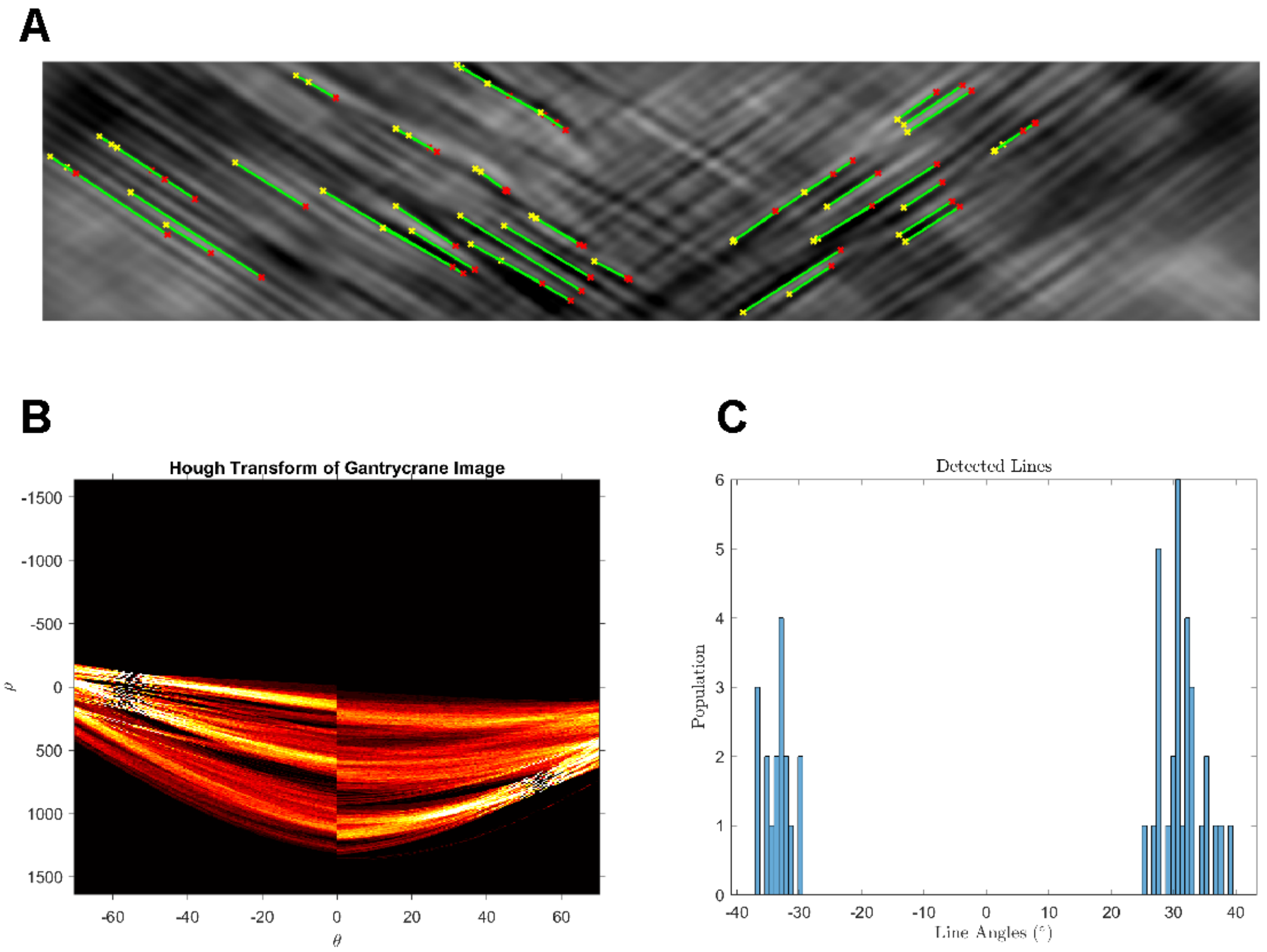

**Fig. S7** **.**
**Hough transform analysis on the FFT-filtered version of the x-ray radiograph image for diamond shocked along [110] at 16 ns**. (**A**) The FFT-filtered version of the x-ray radiograph image with some of the detected lines (green) (**B**) Hough transform matrix ($\rho$, $\theta$). (**C**) The population of the detected line angles.

**Table S1.**
**Determined angles and standard deviations of the detected lines for different shock directions and x-ray delays.**

| Diamond shock direction | X-ray delay [ns] | Averaged angles [degrees] | Standard deviation [degrees] |
|---|---|---|---|
| [100] | 8 | 55 | 3 |
| [100] | 12 | 55 | 2 |
| [100] | 16 | 56 | 3 |
| [110] | 8 | 31 | 3 |
| [110] | 12 | 32 | 3 |
| [110] | 16 | 32 | 3 |
| [111] | 16 | 71 | 3 |

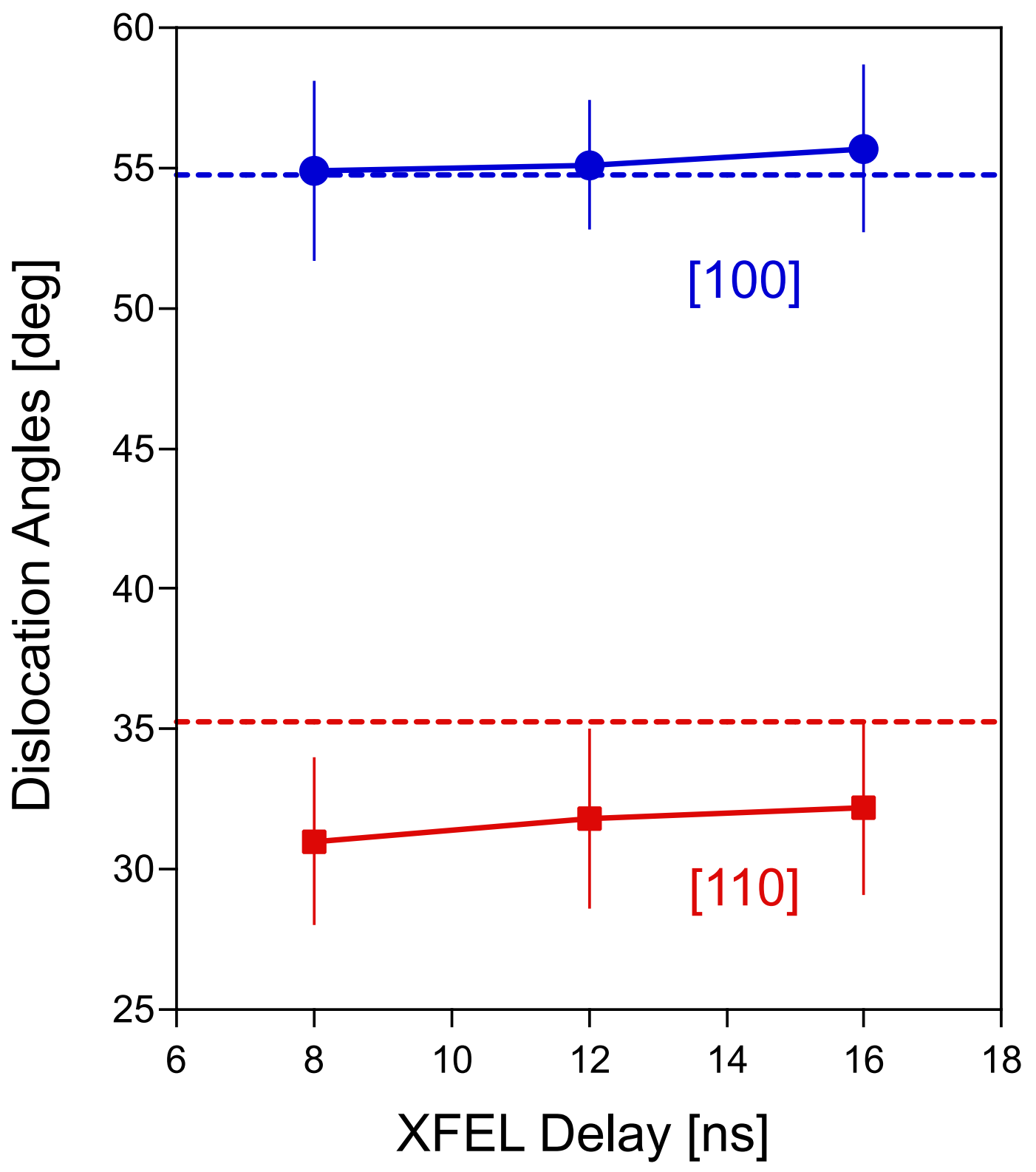

190 **Fig. S8.**

**The measured angles of the detected lines as a factor of the X-ray probe time.** Two different shock directions: [100] (blue circles) and [110] (red squares) are shown. The dashed lines indicate the angle for the undistorted lattice of diamond. Shown error bars represent 1σ of the detected lines. The timing jitter between the drive laser irradiation and the XFEL irradiation is smaller than the size of the shown symbols.

# Supplementary Text

**Diamonds shocked along [111]**

The x-ray radiography images of the diamond shocked along [111] direction are summarized in Fig. S9 to supplement Fig. 2 in the main text that shows the other orientations. While we obtained three images of shocks along the [111] direction, we observe clear stacking fault features only in the image collected at 16 ns. We emphasize that the absence of this image feature at 8 and 12 ns does not indicate the stacking faults suddenly appear between 12 ns and 16 ns, but that there is a shot-to-shot variation in the appearance of the banded structures in our images that complicates their detection for conditions with lower signal-to-noise ratios. Even in the 16 ns shot, the observed lines are lower in intensity than those observed in [100] and [110] orientations, consistent with the strong orientational dependence described in the main text. As described in the main text, we interpret the orientational dependence of the signal contrast for the stacking faults based on the dislocation propagation speed. The speed for [111] shock is not determined here, but since the angle between the stacking faults and diamond surface [(1-11) and (111) in Fig. S9, respectively] is large for the [111] shock case, the dislocation speed for the [111] shock should be slower than those for the other two shock directions. The reason we did not observe any sign of the linear features in the other shots with shorter delays (8 and 12 ns) remains unclear. A possible explanation is that the small difference in the surface conditions of the diamond may have resulted in the shot-by-shot difference of the stacking fault appearance, as the number of pre-existing or shock-induced dislocations at the diamond surface should affect the propagation of the dislocations and the resulting formation of the stacking faults.

A closer look at the 16 ns image for [111] shock shows that some of the stacking faults are left behind the plastic wavefront, indicating that some of the dislocations propagating from the bottom surface of the diamond are stopped or at least slowed down during the propagations. This could be due to the radiation drag or the interaction of the propagating dislocations with pre-existing defects in the diamond, both of which are thought to be suppressed by the stronger driving force for the cases of [100] and [110] shocks. For the [111] shocks, the measured angle between the observed stacking faults and the diamond surface is 71 ± 3°, in agreement with the theoretical value of 71° for an undistorted diamond lattice. Having a large angle between the stacking faults and the diamond surface should result in the stacking fault formation being less drastic, as the dislocation velocity would be slower if the plastic shock wave velocity is similar. Also, the slip deformations are expected to be maximized when the angle between the slip plane and loading direction is 45° (following Schmidt's law) (76).

Note that Germann *et al* performed MD simulations of shocked fcc crystals and found orientational dependence in the dislocation nucleation process (*77*). They observed the motion of partial dislocations with high mobility for shocks along [100] and [110], in contrast to the formation of dislocation loops bounded only by thin stacking-fault ribbons for [111] shocks because the dislocations split into partials. Shock-induced defect evolution and the resulting plasticity have been predicted and observed to depend strongly on the orientation of the crystal as compared to the shock-loading direction. Further investigations on the orientational dependence of the ultrafast defect evolutions by both simulations and experiments would be fruitful.

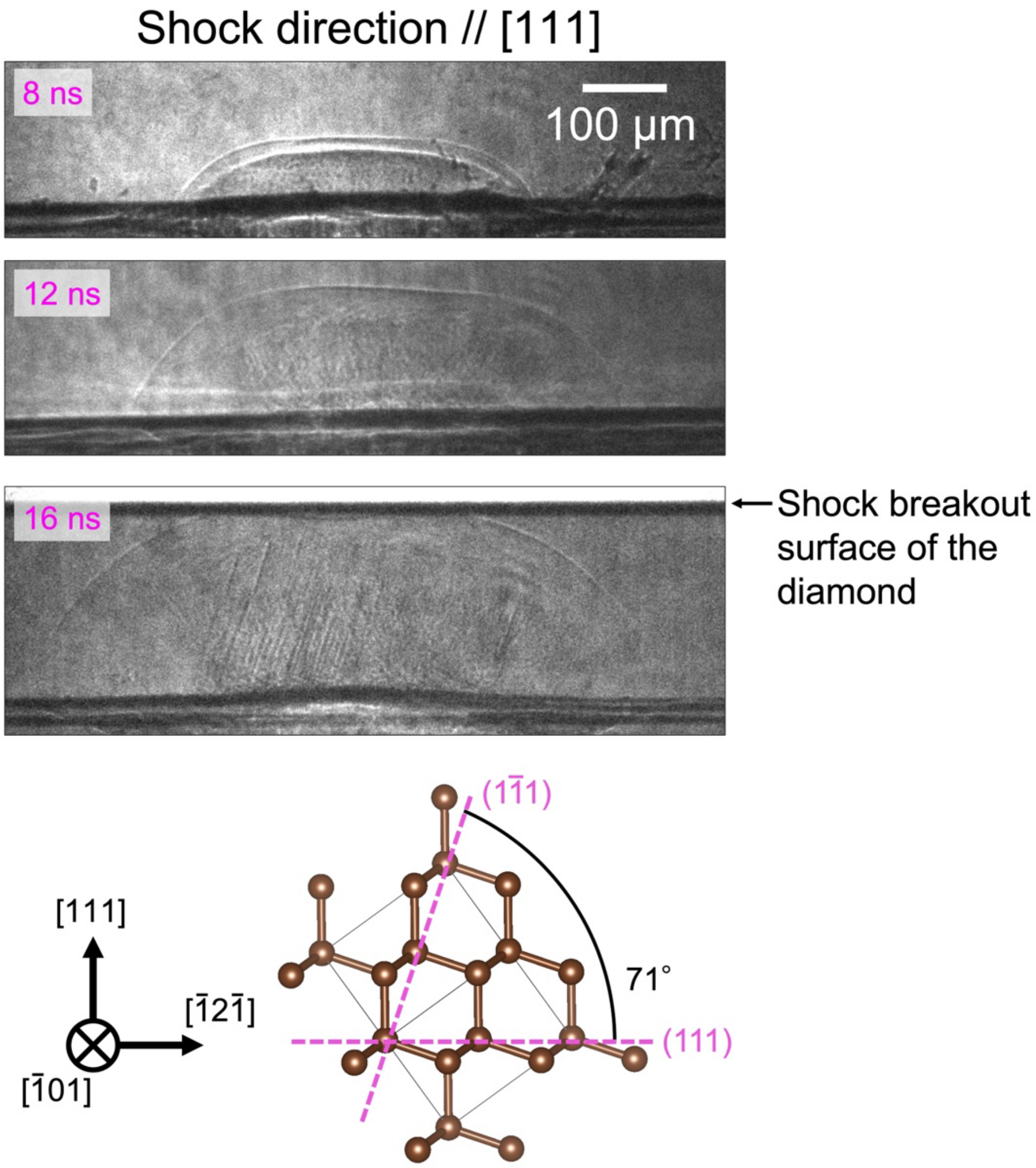

**Fig. S9.**
**X-ray radiography images of diamonds shocked along [111] direction with three different XFEL probe timing.** The diamond lattice structure (*37*) with the corresponding plane orientations is also shown. The elastic wave has already reached the breakout (top) surface of the diamond for the 16 ns shot.

**Probing along different directions**

Since the slip planes are two-dimensional and form along specific directions, the x-ray radiograph images should appear differently when the XFEL probes the crystal from a different direction. Figure S10 summarizes the images of the x-ray radiograph on diamond shocked along [100] but with two different XFEL probe irradiation directions, namely [01-1] and [001] (Fig. S10A and B, respectively). The probing direction was changed by rotating the diamond sample. To make the thickness along the probing direction consistent, the diamond samples were cut along different orientations for the different probing directions. The linear features along the (111) slip planes appear clearly when the diamond is probed along [01-1] while such features are absent when the probe beam is along [001] direction. This indicates that the lines observed at the imaging plane are the results of the x-rays interacting with the shock-induced planar features in the diamond.

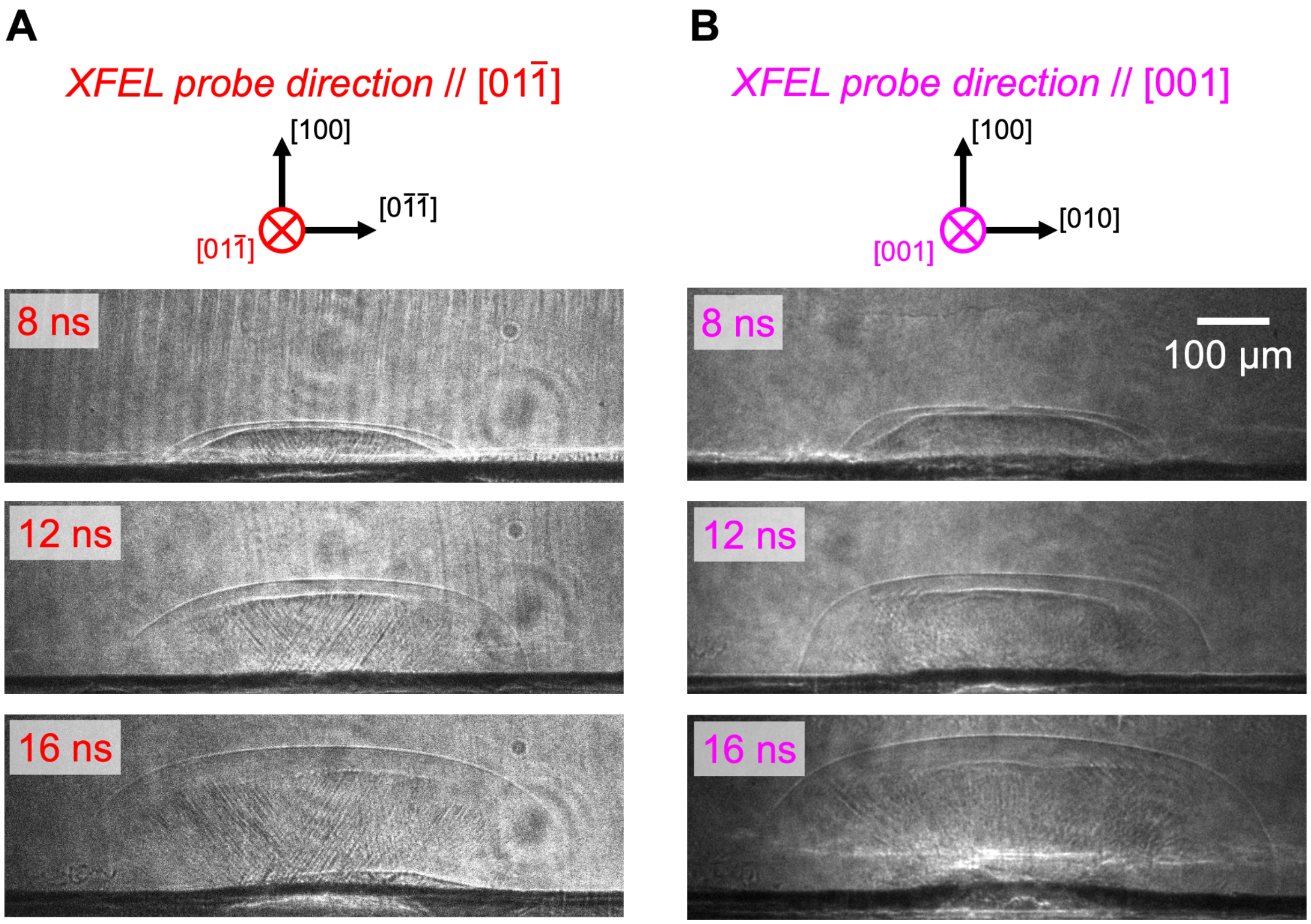

**Fig. S10**

**X-ray radiograph images of diamond shocked along [100] direction but probed along two different XFEL irradiation directions.** The images collected with the XFEL irradiation direction along **(A)** [01-1] and **(B)** [001] of the diamond. The images presented in (A) are identical to the ones in Fig. 2A. The numbers on each image indicate the XFEL probe timing relative to the drive laser irradiation.

**Pre-existing defects in the diamond**

Visible-light microscopy images of the CVD diamond used in this work are shown in Fig. S11. The diamonds shown in these images were collected before the final cuttings and polishings, and thus are larger than the actual sample size used for the x-ray radiograph experiments. The pattern seen in Fig. S11B shows the cross-polarized microscopy image whose features indicate strain caused by pre-existing defects. This suggests that the diamonds used in this work had bundles of dislocations threading out of the growth plate which are typical defects observed in CVD-grown diamonds. Note that the observed transonic dislocations could be both pre-existing and shock-induced (newly created) dislocations at the surface or sub-surface of the diamond.

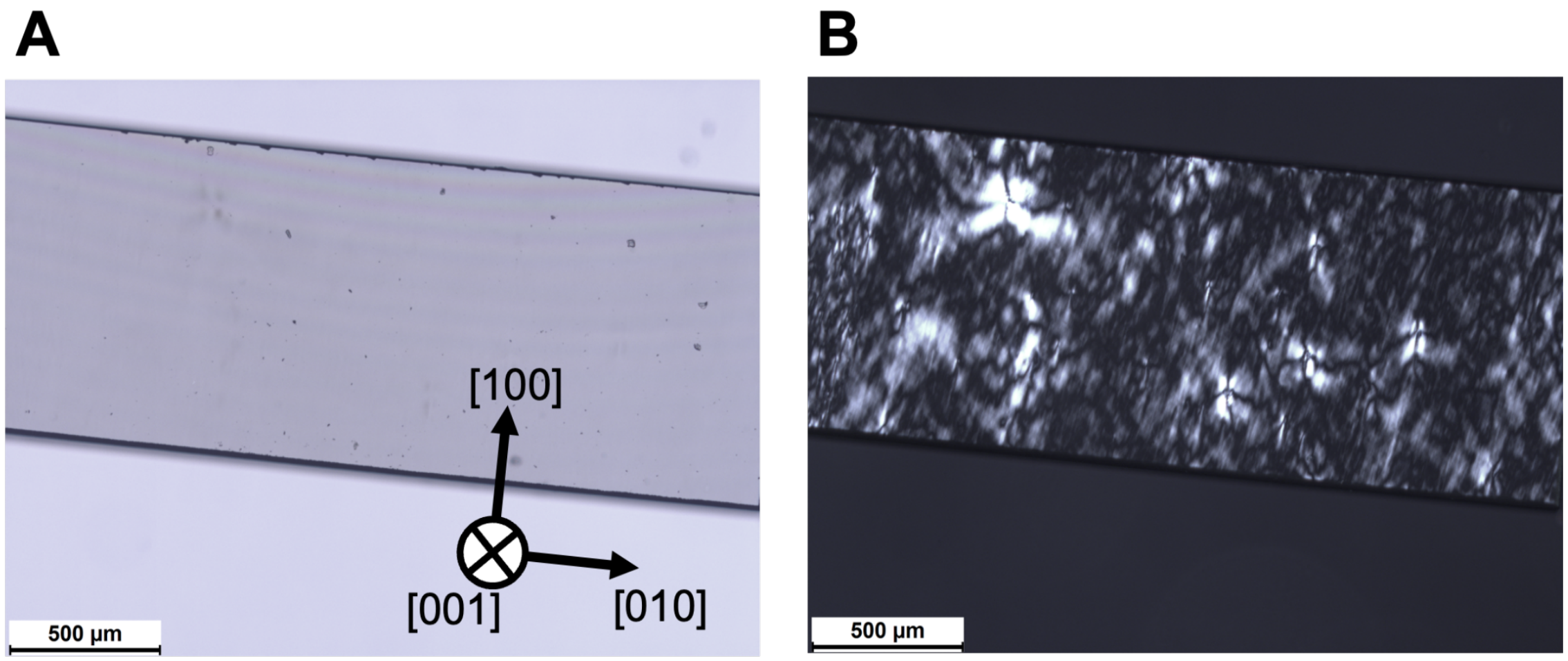

**Fig. S11.**
**Optical microscopy images of the CVD type IIa diamond used in this work.** (**A**) unpolarized and (**B**) polarized images obtained by using an optical microscope with cross-polarizers. The diamond orientations are indicated in A.

**Shock dynamics at delays beyond 16 ns**

X-ray radiography images of diamond shocked along [100] and probed along [011] are summarized in Fig. S12 over a wider range of x-ray probe timings. After the shock waves reach the breakout (top) surface of the diamond at a ~20 ns delay, the breakout surface freely expands into the vacuum. In the 300-ns and 500-ns shots, many small particles are formed as the diamond is shock blasted into the vacuum, indicating the occurrence of grain refining during the pressure release that facilitates the subsequent fracture. As the peak stresses achieved in this study are <20% of the shock melting point of diamond (P ~ 680 GPa and T > 7000 K) (*45*,*46,78*), all of the observed image features should remain in the solid state.

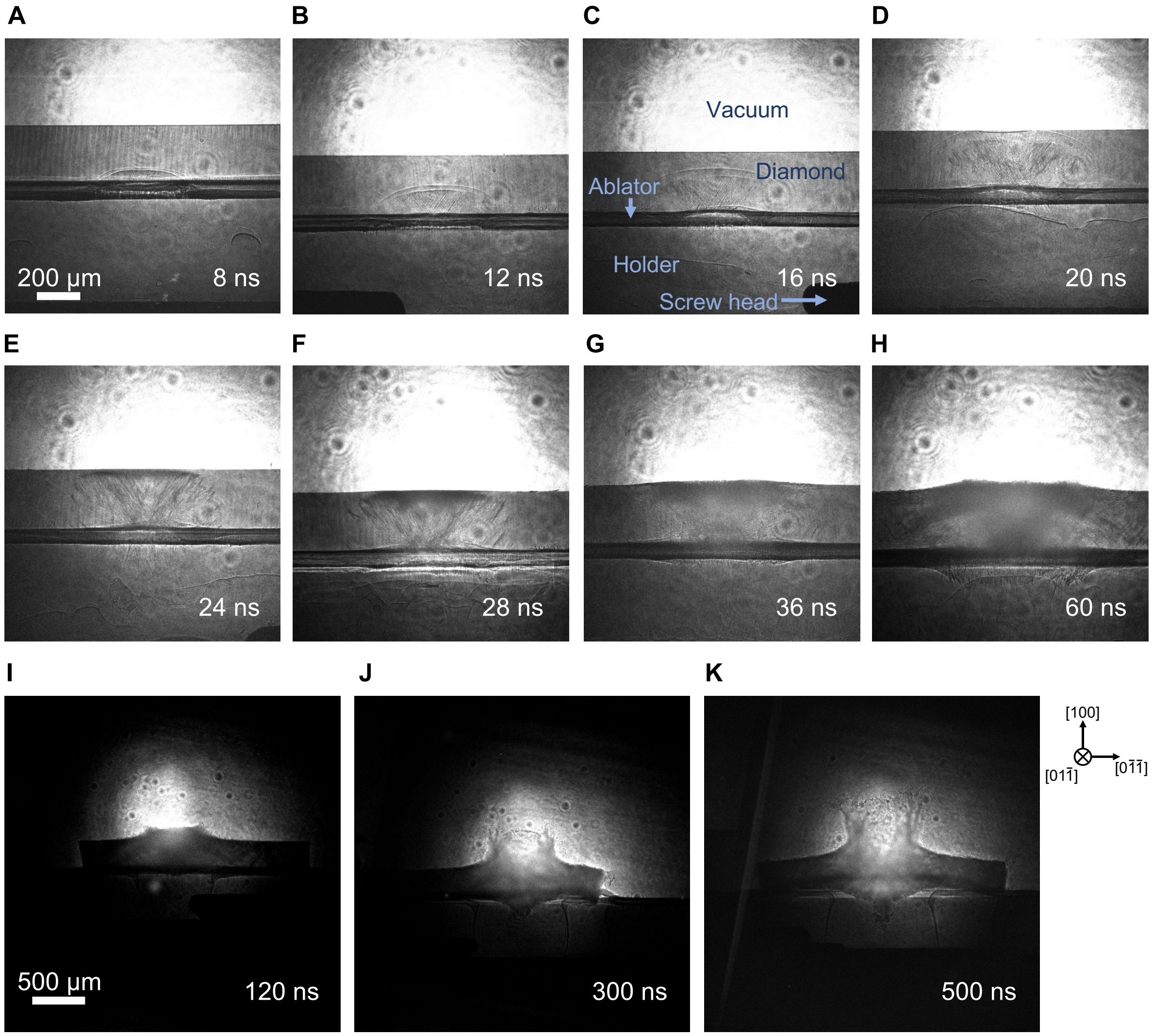

**Fig. S12.**

**The x-ray radiography images of diamond shocked along [100] at x-ray probe delays up to 500 ns.** The XFEL beam was irradiated along the diamond [01-1] direction. A-C are identical to the images shown in Fig. 2A but uncropped. The magnification of the microscope used during the read-out process is different between A-H (10x) and I-K (4x), as the expansion of the shock-blasted diamond became significant at longer delays (120, 300, and 500 ns) and the observed features do not fit into the field-of-view of the 10x magnification image. The shock target is placed on the holder as denoted in C. The holder is a plastic ring shaped like a washer, and the laser goes through the hole of the ring, hitting the ablator directly.